# Gravitational wave signals from finite size effects in spinning binary inspirals including parity violating constituents

**Beka Modrekiladze**

*Department of Physics, Carnegie Mellon University,
PA 15213, U.S.A.*

*E-mail:* bekam@cmu.edu

ABSTRACT: We generalize the world line EFT formalism to account for parity violating finite size effects. Results are presented for potentials and radiating moments of a binary inspiral for the parity conserving sector, and agreement is found with, previous calculations. Furthermore, we generate new results in this sector, calculating the current quadrupole moment induced by finite size gravitomagnetic effects. We also present novel results for parity violating sources, which might be due to beyond standard model physics, and show that they generate GW signals with the unique signature that the current-moment appears at 0.5PN order earlier relative to the mass-moment in the PN expansion. Parity violation also induces a new type of potential, which is proportional to the $\mathbf{S} \cdot \mathbf{r}$. Finally, we present new results for the dissipative force for parity violating constituents, which leads to the curious signature of a force normal to the orbit.

https://doi.org/10.1007/JHEP03(2023)019

## Contents

## 1 Introduction

With the detection of gravitational waves, we have a new way to search for exotic matter beyond the Standard Model (SM). We can learn about the properties of matter through the finite size effects captured in the GW signal. Several such possible exotic compact objects have been proposed, including boson stars [5], gravastars [6], quark stars [7] and axion stars [8, 9].

Of particular interest is the possibility of parity violating constituents since there is no reason to expect exotic stars to be parity conserving, given that parity is maximally violated in the SM. In this paper, we show that effects from the parity violating terms have a unique signature in the GW signal.

With the third generation of gravitational wave observatories, it should be possible to detect the finite size effects in the GWs signal. However, since matched-filtering uses a bank of template waveforms [10, 11], we will possible miss signals for which we do not have the corresponding waveforms in the catalog, due to the lower signal-to-noise-ratio. Therefore, having a formalism, which includes finite size effects, is essential.

In section 2, we review the general formalism of finite size effects in the effective field theory approach. In the next section, section 3, we calculate their effects on potentials and radiation moments and compare to the previously obtained results. Section 4 presents new results for parity violating finite size effects. We discover that the current quadrupole appears at 0.5 Post-Newtonian(PN) order earlier than the mass quadrupole, which is a unique feature for the parity violating effects. We also generalize the problem by adding the spin and discover another unique property — a potential, which is proportional to the $\mathbf{S}\cdot\mathbf{r}$.

Finite size effects introduce additional degrees of freedom which lead to absorptive effects. We calculate, in section 4.4, the resulting parity violating dissipative force, which is directed normal to the orbit and induces oscillations outside of the plane of inspiral. The corresponding induced behavior of the orbit is explored qualitatively. Finally, in section 6 we conclude and discuss further prospects.

## 2 EFT

For completeness, we give a rapid review of the formalism developed in [1, 2]. We start by constructing the relevant action for an isolated compact body. At the level of a point particle, we parameterize the particle's world line via $x^\mu(\lambda)$. Then, the action for the point particle (in units $c = \hbar = 1$) is given by

$$S_0 = -M \int d\tau - 2M_{pl}^2 \int d^4x \sqrt{g} R. \tag{2.1}$$

Now we would like to generalize the point particle action to account for finite size effects. This is done by writing down all terms in the action that are consistent with general coordinate invariance and world line reparameterization invariance. Finite size effects are accounted for by including the lowest dimension terms, which are also even in parity. There are two such terms bi-linear in the Weyl tensor,

$$S_{E+B} = \int d\tau (C_E E_{\mu\nu} E^{\mu\nu} + C_B B_{\mu\nu} B^{\mu\nu}), \tag{2.2}$$

where coefficients $C_E$ and $C_B$ are proportional to the electric ($\lambda$) and magnetic ($\sigma$) tidal deformabilities, correspondingly. Higher dimension terms are suppressed in the low frequency expansion. We add a gauge fixing term $S_{GF}$, which has the following form

$$S_{GF} = M_{pl}^2 \int d^4x \sqrt{g} \Gamma_\mu \Gamma^\mu \tag{2.3}$$

with $\Gamma_\mu = \partial_\alpha H_\mu^\alpha - \frac{1}{2}\partial_\mu H_\alpha^\alpha$.[1]

We next consider the interaction between two compact objects in the PN expansion. We decompose the graviton as

$$g_{\mu\nu} = \eta_{\mu\nu} + \frac{h_{\mu\nu}(x)}{M_{pl}} = \eta_{\mu\nu} + \frac{\bar{h}_{\mu\nu}(x)}{M_{pl}} + \frac{H_{\mu\nu}(x)}{M_{pl}}, \tag{2.4}$$

where $H_{\mu\nu}(x)$ represents the potential gravitons and $\bar{h}_{\mu\nu}(x)$ describes radiation gravitons, and their momenta scale as $(\frac{1}{r}, \frac{\mathbf{v}}{r})$ and $(\frac{|\mathbf{v}|}{r}, \frac{\mathbf{v}}{r})$, respectively. The relevant interaction vertices

---

[1]It provides us following instantaneous propagator for the potential graviton modes,

$$< H_{\mu\nu}^\mathbf{k}(x^0) H_{\alpha\beta}^\mathbf{q}(0) > = -(2\pi)^3 \delta^3(\mathbf{k}+\mathbf{q})\frac{i}{\mathbf{k}^2}\delta(x_0) P_{\mu\nu,\alpha\beta},$$

where $P_{\mu\nu,\alpha\beta} = \frac{1}{2}\left[\eta_{\mu\alpha}\eta_{\nu\beta} + \eta_{\mu\beta}\eta_{\nu\alpha} - \frac{2}{d-2}\eta_{\mu\nu}\eta_{\alpha\beta}\right]$ with $d$ the spacetime dimension and $\mathbf{k}^2$ is 3-momentum square, choosing $k_0 = 0$.

derived from the $-M \int d\tau$ are:

$$L_{v^0} = -M \frac{1}{2} \frac{H_{00}}{M_{pl}} \quad O(1)$$

$$L_{v^1} = -M v^i \frac{H_{0i}}{M_{pl}} \quad O(v)$$

$$L_{v^2} = -M \frac{1}{2} v^i v^j \frac{H_{ij}}{M_{pl}} - M \frac{1}{4} v^2 \frac{H_{00}}{M_{pl}} + M \frac{1}{8} \frac{H_{00}}{M_{pl}} \frac{H_{00}}{M_{pl}} \quad O(v^2)$$

$$\ldots$$

In a similar way, we are interested in the scaling of vertex factors in the $E$ and $B$ parts of the Weyl tensor. The linearized version of the electric part of the Weyl tensor has the following form

$$E_{ij} = \frac{1}{2M_{pl}} \Big( (h_{i0,0j} - h_{00,ij}) - (h_{ij,00} - h_{j0,0i}) \Big), \tag{2.5}$$

where each term has the definite scaling in the velocity. For potential modes, temporal and spatial derivatives scale differently, $\partial_0 \sim \frac{v}{r}$ and $\partial_i \sim \frac{1}{r}$. Thus, the leading order terms in the $E$ part of action are:

$$C_E E_{ij} E^{ij} = \frac{1}{4M_{pl}^2} \Big( C_E h_{00,ij} h_{00,ij} \quad O(\lambda)$$

$$-4 C_E h_{i0,0j} h_{00,ij} \Big) \quad O(\lambda v)$$

$$\ldots$$

Following the same procedure for the B part, we write down the linearized version of the magnetic part of the Weyl tensor,

$$B_{ij} = \frac{\epsilon_{imn}}{2M_{pl}} (h_{n0,jm} + h_{jm,0n}). \tag{2.6}$$

Which gives the following homogeneous scaling in the velocity,

$$C_B B_{ij} B^{ij} = \frac{1}{4M_{pl}^2} \epsilon_{imn} \epsilon_{ipq} \Big( C_B h_{n0,jm} h_{q0,jp} \quad O(\sigma)$$

$$+ 2 C_B h_{n0,jm} h_{jp,0q} \Big) \quad O(\sigma v)$$

$$\ldots$$

Since we have the dimensionful coupling in the gravity, we can conclude that

$$C_{E/B} \sim R^3 (R M_{pl})^n, \tag{2.7}$$

where $R$ is the typical length scale of the compact object. However, $C_{E/B}$ are proportional to the tidal deformabilities, which scale as $R^5$. From which, due to eq. (2.7), follows that $n$ needs to be equal to 2. We now have machinery for computing potentials and moments, to which the following section will be dedicated. For further review of the subject you can see [12–16].

– 3 –

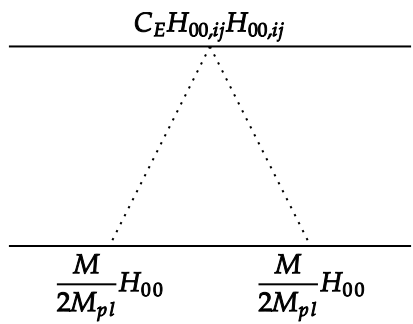

**Figure 1**. Leading Order $E$-potential.

## 3 Parity conserving sources

We first explore parity conserving effects and compare them to the known results. In so doing we derive the relation between $C_E, C_B$ and the standard definitions of Love numbers.

### 3.1 Gravitoelectric effects

To calculate potentials, we integrate out the potential modes. The leading order diagram for gravitoelectric potential, shown in figure 1, scales as $O(\lambda v^0)$, and it is given by

$$-iTV^{O(\lambda)} =$$
$$(-i)^3 \frac{1}{2!} \left(\frac{M_1}{2M_{pl}}\right)^2 \frac{C_E}{4M_{pl}^2} \int dt_1 dt_2 < H_{00}(x_1(t_1))H_{00}(x_1(t_1'))H_{00,ij}(x_2(t_2))H_{00,ij}(x_2(t_2)) > . \tag{3.1}$$

This gives us the following expression for the leading order potential of finite size effect by gravito-electric tidal interactions

$$V = -\frac{G_N M_1^2}{r}\left(\frac{3}{16\pi}\right)\frac{C_E}{M_{pl}^2 r^5} = -\frac{6G_N^2 M_1^2}{r^6}C_E. \tag{3.2}$$

We can now use this result to relate our coefficient to the tidal Love number. We compare to the results of [3], which provides us with the potential, with gravitoelectric effects, up to the order $O(\lambda v^2)$

$$V[z^i] = \frac{\mu M}{r}\left(1 + \frac{\Lambda}{r^5}\right) + O(c^{-2}\lambda) + O(c^{-4}) + O(\lambda^2). \tag{3.3}$$

The only $O(\lambda)$ term is: $\frac{\mu M}{r^6}\Lambda$, with $\Lambda = \frac{3M_1}{2M_2}\lambda$. By convention, the static Love number $k_2$ is related to the tidal deformability parameter in the following way $\lambda = \frac{2}{3}k_2 R^5$. Comparing it to the corresponding potential eq. (3.2) of our results, we can fix the value of $C_E$.

$$C_E = \frac{16\pi}{3}M_{pl}^2 R^5 k_2. \tag{3.4}$$

Let us now calculate the mass multipole moments due to gravitoelectric tidal effects. This will allow us to check the consistency of the relation eq. (3.4). The leading order contribution to the mass quadrupole comes at $O(\lambda v^0)$, with one radiating graviton, as shown in figure 2. By definition, the energy-momentum tensor is the variation of the matter action $S$ by the metric

$$\int d^4 x T_{\mu\nu} = -2 M_{pl} \frac{\delta S}{\delta h_{\mu\nu}}. \tag{3.5}$$

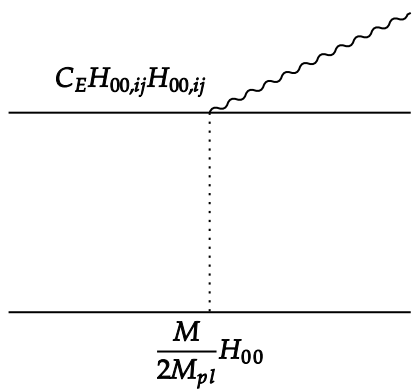

**Figure 2**. Leading Order Mass Quadrupole.

Thus, for the leading order diagram, we have

$$T_{00}(q) =$$
$$<0| -2 M_{pl} \frac{\delta}{\delta h_{00}} \int dt_1 dt_2 (-M_1) \frac{1}{2 M_{pl}} \frac{H_{00}(x,t)}{M_{pl}} C_E \frac{2}{4 M_{pl}} H_{00,ij}(x_2,t_2) \bar{h}_{00,ij}(x_2,t_2) |q>. \tag{3.6}$$

Which gives us the 00 component of energy-momentum tensor in momentum space,

$$T_{00}(q) = \frac{M_1 C_E}{2 M_{pl}^2} \frac{1}{2} \int dt \frac{3 n^{<ij>}}{4 \pi r^3} (-q_i q_j) e^{-iqx_2}, \tag{3.7}$$

where $n^{<ij>} = n^i n^j - \frac{1}{3} \delta^{ij}$. Note that the Fourier transform of energy-momentum tensor has the following form

$$T^{\mu\nu}(x^0, \mathbf{q}) = \int d^3 x T^{\mu\nu}(x) e^{iqx} = \int d^3 x T(x^0, \mathbf{x}) \sum_{n=0}^{\infty} \frac{(iqx)^n}{n!}. \tag{3.8}$$

We can obtain second moments of the energy-momentum tensor in the coordinate space by

$$-\frac{1}{2!} \int d^3 x T^{\mu\nu}(x) x_i x_j = \frac{\partial}{\partial q_i} \frac{\partial}{\partial q_j} T^{\mu\nu}(x^0, q)|_{q=0}, \tag{3.9}$$

which for our result, eq. (3.7), gives

$$2! \frac{M_1 C_E}{4 M_{pl}^2} \frac{3 n^{<ij>}}{4 \pi r^3} = \int d^3 x T^{00} x^i x^j. \tag{3.10}$$

Then, the quadrupole moment is given by

$$Q_{ij} = \frac{M_1 C_E}{2 M_{pl}^2} \frac{3 n^{<ij>}}{4 \pi r^3}. \tag{3.11}$$

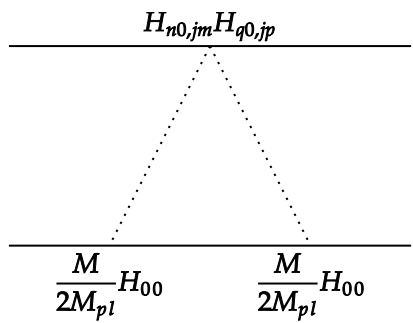

**Figure 3**. Leading Order B-potential.

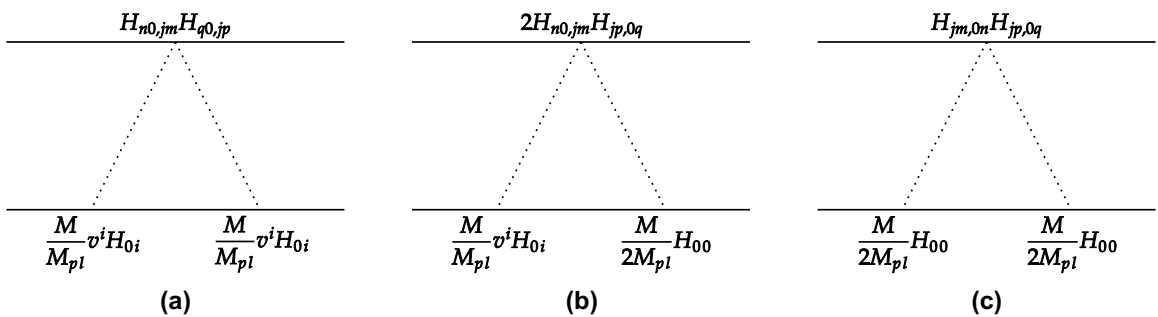

**Figure 4**. Next to Leading Order B-potentials.

By comparing to the terms of corresponding order in the results of [3],

$$Q_2^{ij}(t) = \lambda \frac{3\chi_1 M}{r^3} n^{<ij>} + \lambda \frac{1}{c^2} \frac{3\chi_1 M}{r^3} \left[ \left( 2v^2 - \frac{5\chi_2^2}{2}\dot{r}^2 - \frac{6-\chi_2}{2}\frac{M}{r} \right) n^{<ij>} + v^{<ij>} - (3-\chi_2^2)\dot{r} n^{<i} v^{j>} \right]$$
$$+ O(\lambda c^{-4}) + O(\lambda^2),$$

where $\chi_1 = \frac{M1}{M}$ and $\chi_2 = \frac{M2}{M}$, we get the same value for our coefficient, $C_E = \frac{16\pi}{3} M_{pl}^2 R^5 k_2$.

## 3.2 Gravitomagnetic effects

To generate the leading order potentials in $C_B$, we consider diagrams with one $B^2$ vertex and two mass insertions. We have the only one possible diagram figure 3 at the leading order in the PN expansion.

Figure 3, is proportional to $<H_{n0}H_{00}>$, which is zero, since in our gauge

$$<H_{n0}H_{00}> \sim P_{n0,00} = 0. \tag{3.12}$$

At the next order, we have three possible diagrams, which give non-zero contractions.

The sum of those diagrams gives

$$V^{O(C_B v^2)} = -\frac{4 C_B M_2^2}{(32\pi M_{pl})^2 r^6} \left( \frac{3}{2} \right) \left( (\mathbf{v} \cdot \mathbf{n})^2 - v^2 \right). \tag{3.13}$$

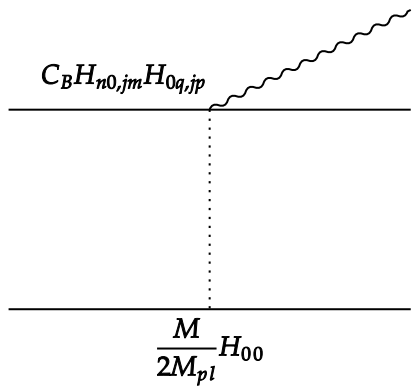

**Figure 5**. Leading Order Current Quadrupole.

We can use this result to relate our coefficient $C_B$ to the quadrupolar gravitomagnetic tidal Love number, $k_2^{mag}$. We compare to the results of [4] which provides us with the potential, with gravitomagnetic effects, up to order $O(\sigma v^2)$,[2]

$$V = \frac{GM\mu}{r} + \frac{12G^2 m_2^2 \sigma_1}{r^6}\left(v^2 - (\mathbf{v}\cdot\mathbf{n})^2\right). \tag{3.14}$$

Comparing to our result, eq. (3.13), we deduce that

$$C_B = 2\sigma_1. \tag{3.15}$$

In the case of gravitomagnetic effects, we calculate the induced current quadrupole, since it's the leading order multipole moment for $B^2$. At that order the only possible diagram is shown in figure 5 and is proportional to $<H_{00}H_{n0}>$, which is zero. Therefore, there is no contribution at the leading order.[3] The leading order contribution will be at order $O(v\sigma)$ and is given by diagrams shown in figure 6 below.

The sum of diagrams in figure 6 gives

$$T_{0n}(q) = \frac{C_B M_2}{4M_{pl}^2}\frac{2}{4\pi r^3}\int dt (v_n q^2 - (v\cdot q)q_n)e^{-iqx_2}. \tag{3.16}$$

From eq. (3.16), we can read of following moments:

$$\int d^3 x T_{n0}(x^0,x) x^2 = \frac{C_B M_2}{4M_{pl}^2} v_n, \tag{3.17}$$

$$\int d^3 x T_{n0}(x^0,x) x_i x_n = \frac{C_B M_2}{4M_{pl}^2} v_i. \tag{3.18}$$

For the current quadrupole, we need $\left(\int d^3 x T_{n0}(x^0,x) x_i x_j\right)_{TF}$ and we can obtain it from the only diagram which is left at the order $O(v\sigma)$. We find

$$T_{0n}(q) = \frac{C_B M_2}{2M_{pl}^2}\frac{3}{4\pi r^3}\int dt (v_n n^{<jp>} q_j q_p - (v\cdot q)q_j n^{jn})e^{-iqx_2}, \tag{3.19}$$

---

[2]By convention $\sigma = \frac{1}{2c^2} k_2^{mag} M R^5$.

[3]In general, it can be shown that diagrams that scale by even power of velocity will be zero.

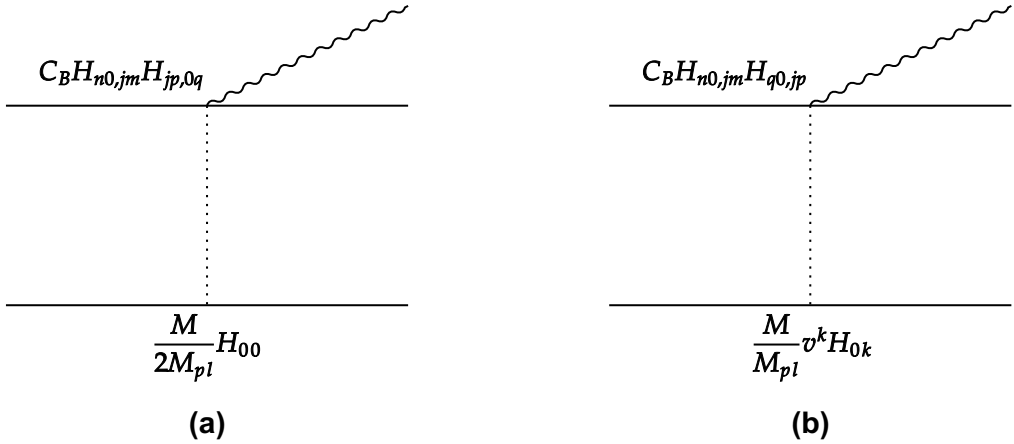

**Figure 6**. Next to Leading Order Current Quadrupoles.

from which we can extract,

$$\int d^3x T_{n0}(x^0,x)x_i x_j = \frac{C_B M_2}{2M_{pl}^2}\frac{3}{4\pi r^3}(v_n n^{<ij>} - v_i n^{<jn>}) \equiv A_{nij}\frac{C_B M_2}{2M_{pl}^2}\frac{3}{4\pi r^3}. \qquad (3.20)$$

Therefore, the current multipole, induced by $C_B$, is given by

$$j_{ik} = \epsilon^{kjn}\int d^3x T_{n0}(x^0,x)x_i x_j \equiv \epsilon^{kjn} A_{nij}\frac{C_B M_2}{2M_{pl}^2}\frac{3}{4\pi r^3}. \qquad (3.21)$$

## 4 Parity violating sources

Now we generalize the action by adding an odd-parity finite size effects term, $S_{EB} = \int d\tau C E_{\mu\nu} B^{\mu\nu}$. To the best of our knowledge, this is a new result. To power count, we group terms by their scaling in velocity.

$$CE_{ij}B^{ij} = \frac{\epsilon_{imn}}{4M_{pl}^2}\Big(-Ch_{00,ij}h_{n0,jm} \quad O(C)$$

$$+C((h_{i0,0j} + h_{j0,0i})h_{n0,jm} - h_{00,ij}h_{jm,0n})\Big) \quad O(Cv)$$

$$\cdots$$

### 4.1 Current quadrupole

To calculate moments, we use the same approach as we used for the parity conserving sector. With one radiating graviton, we have only one diagram at the leading order which is shown in figure 7 below.

From figure 7, we can extract $T_{0n}$.

$$T_{0n}(q) = 2M_{pl}\frac{\delta}{\delta \bar{h}_{0n}} <0|S|q> = 2M_{pl}\frac{M_1}{2M_{pl}}\frac{C}{4M_{pl}^2}\epsilon_{imn}\frac{\delta}{\delta \bar{h}_{0n}} <0|\int d^3x H_{00,ij}H_{00}\bar{h}_{0n,jm}|q>$$

$$= \frac{CM_1}{4M_{pl}^2}\epsilon_{imn}\frac{n^{<ij>}}{4\pi r^3}q_j q_m e^{-iqr}. \qquad (4.1)$$

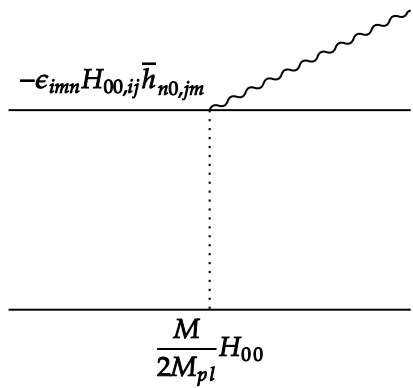

**Figure 7**. Leading Order EB-Current Quadrupole.

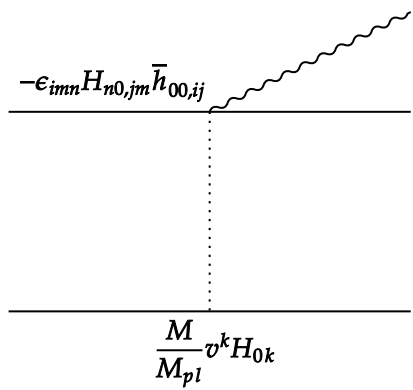

**Figure 8**. Leading order EB-Mass Quadrupole.

Thus, we can extract the current quadrupole

$$\int d^3x T_{n0}(x) x^i x^m = \frac{M_1 C}{4M_{pl}^2} \epsilon_{imn} \frac{n^{<ij>}}{4\pi r^3}. \tag{4.2}$$

### 4.2 Mass quadrupole

To extract the mass quadrupole, we have also one possible diagram at the leading order which is shown in figure 8 below,

which provides the following components of energy-momentum tensor

$$T_{00}(q) = 2M_{pl}\frac{\delta}{\delta \bar{h}_{00}} < 0|S|q> = 2M_{pl}\frac{M_1 v^k}{M_{pl}}\frac{C}{4M_{pl}^2}\epsilon_{imn}\frac{\delta}{\delta \bar{h}_{00}} < 0|\int d^3x H_{n0,jm} H_{0k} \bar{h}_{00,ij}|q>, \tag{4.3}$$

$$T_{00}(q) = \frac{CM_1}{4M_{pl}^2} v_n \epsilon_{imn} \frac{n^{<jm>}}{4\pi r^3} q_i q_j e^{-iqr}. \tag{4.4}$$

And we extract mass quadrupole,

$$Q_{ij} = \frac{CM_1}{4M_{pl}^2} v_n \epsilon_{imn} \frac{n^{<jm>}}{4\pi r^3}. \tag{4.5}$$

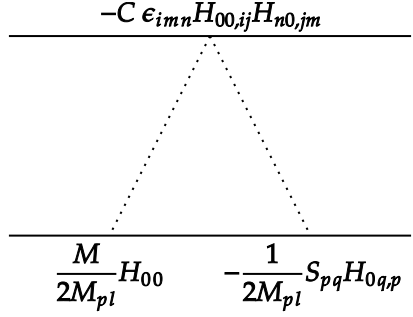

**Figure 9**. Leading Order Spin+EB Potential.

We find the mass quadrupole, induced by parity violating effects, is (+0.5PN) relative to the current quadrupole (4.2). Thus, the current quadrupole will appear earlier than the mass quadrupole effects in the PN order, which is not typical for a partiy conserving case.

### 4.3 Spin

The second unique feature appears in the potential if we take into account the spin of the binary partner. Also, without the spin terms, potentials by the parity violating tidal terms vanish. Since $S$ is a conjugate momentum, we use a Routhian description of the system, where the Routhian $R$ has a role of Hamiltonian for the spin degrees of freedom and the leading order Lagrangian for the world line is [20]

$$R = -\frac{1}{2M_{pl}}(S_{ji}h_0^{i,j} + S_{j0}h_0^{0,j}). \tag{4.6}$$

To the leading order, only the first term of eq. (4.6) contributes, figure 9.

The diagram in figure 9 gives us the following results for the potential,

$$V = -\frac{1}{2M_{pl}}\frac{M}{2M_{pl}}\frac{C\epsilon_{imn}}{4M_{pl}^2}S_{ab} <H_{0b,a}H_{n0,jm}><H_{00}H_{00,ij}>. \tag{4.7}$$

Calculating the correlators,

$$\int_k\int_l\int_p\int_q e^{ikx_1}e^{ilx_1}e^{ipx_2}e^{iqx_1}k_ap_jp_m <H_{0b}^k H_{n0}^p> q_iq_j <H_{00}^l H_{00}^q> = \frac{\delta_{bn}}{4}\left(\partial_a\frac{3n^{<jm>}}{4\pi r^3}\right)\frac{3n^{<ij>}}{4\pi r^3}, \tag{4.8}$$

which gives us the expression for the potential,

$$V = \frac{9MC}{512\pi^2 M_{pl}^4}\frac{\mathbf{S}\cdot\mathbf{r}}{r^8}. \tag{4.9}$$

This potential results in novel orbital evolution. To see this, note that the force,

$$\mathbf{F} = -\nabla V = -\frac{9MC}{512\pi^2 M_{pl}^4}\frac{\mathbf{S}}{r^8} + \frac{9MC}{512\pi^2 M_{pl}^4}\frac{8(\mathbf{S}\cdot\mathbf{r})}{r^9}\hat{r}, \tag{4.10}$$

contains a piece, proportional to the direction of spin. And when spin directed out of the orbital plane, $\mathbf{S}\cdot\mathbf{r} = 0$, it leads to the following equation of motion in the normal direction.

$$M_1\ddot{z} = -\frac{9M_1C}{512\pi^2 M_{pl}^4}\frac{S_z}{r^8} - \frac{1}{32\pi M_{pl}^2}\frac{M_1M_2}{r^2}\frac{z}{r}. \tag{4.11}$$

This generates oscillations perpendicular to the orbital plane. In otder to analyze eq. (4.11) numerically, we rewrite it as[4]

$$\ddot{z} = -\frac{A}{(R^2+z^2)^4} - \frac{Bz}{(R^2+z^2)^{3/2}}. \tag{4.12}$$

Since we consider the inspiral phase, $z \ll R$, we expand eq. (4.12) in that limit. At the linear order, we have,

$$\ddot{z} + \frac{B}{R^3}\left(z + \frac{A}{BR^5}\right) = 0. \tag{4.13}$$

The constant force shifts the equilibrium point from $z = 0$ to $z_0 = 18\frac{CS_z}{G}$ and the frequency of oscillations remains the same, $\omega_z = \sqrt{\frac{GM_2}{R^3}}$. This frequency coincides with the orbital frequency, which at first glance may seem surprising. However, the reason is that, at the linear order, the Newtonian potential acts as a restoring force, and the constant force effectively only shifts the equilibrium.

Interestingly, linear appearance of $\mathbf{S} \cdot \mathbf{r}$ in potential to be only due to the parity violating effects, holds true in the case of short-distance modifications to general relativity, via higher curvature terms in fundamental gravity, as well. Considering the effective Lagrangian proposed in [36]

$$S_{eff} = \frac{1}{16\pi G} \int d^4x \sqrt{-g} \left( R - \frac{1}{2}\Gamma^\mu \Gamma_\mu + \frac{\mathcal{C}^2}{\Lambda^6} + \frac{\mathcal{C}\tilde{\mathcal{C}}}{\Lambda_-^6} + \frac{\tilde{\mathcal{C}}^2}{\tilde{\Lambda}^6} \right), \tag{4.14}$$

where $R$ is the Ricci scalar, $\Gamma^\mu \equiv g^{\nu\rho}\Gamma^\mu_{\nu\rho}$ enters the gauge fixing term and

$$\mathcal{C} \equiv R_{\alpha\beta\gamma\delta} R^{\alpha\beta\gamma\delta}, \tag{4.15}$$

$$\tilde{\mathcal{C}} \equiv R_{\alpha\beta\gamma\delta} \epsilon^{\alpha\beta\mu\nu} R_{\mu\nu\rho\sigma} g^{\gamma\rho} g^{\delta\sigma}. \tag{4.16}$$

In [37], lowest order linear-in-spin contributions to the potential were calculated for $\mathcal{C}^2$, $\mathcal{C}\tilde{\mathcal{C}}$ and $\tilde{\mathcal{C}}^2$ terms. Similarly, only the parity violating term, $\mathcal{C}\tilde{\mathcal{C}}$, results to the potential, linear to $\mathbf{S} \cdot \mathbf{r}$

$$V_{\Lambda_- S} = -4608 \frac{G^3 M_1^2 M_2}{r^5 (\Lambda_- r)^6} \mathbf{S} \cdot \mathbf{r}. \tag{4.17}$$

While this potential has a similar form as eq. (4.9), it is relatively suppressed by $M_{pl}^{-2}$. This is due to the fact that it was obtained from the higher curvature terms in eq. (4.14) relative to the Einstein-Hilbert action. One might expect that having $V \propto \mathbf{S} \cdot \mathbf{r}$, only in the case of parity violation, to be ubiquitous for the next orders of possible UV completions of general relativity as well.

### 4.4 Absorption

By introducing finite size effects, we allow our compact objects to deform, leading to dissipation. To account for dissipation, we have to introduce the extra degrees of freedom that live on the world-line. We introduce two operators on the world-line $Q_{\mu\nu}(\tau)$ and $M_{\mu\nu}(\tau)$ which couple to $E$ and $B$ fields[5][17–21],

$$S_{dis} = \int d\tau (Q_{ab}E^{ab} + M_{ab}B^{ab}). \tag{4.18}$$

---
[4]Where $A = \frac{9C}{512\pi^2 M_{pl}^4} S_z$ and $B = \frac{M_2}{32\pi M_{pl}^2}$.

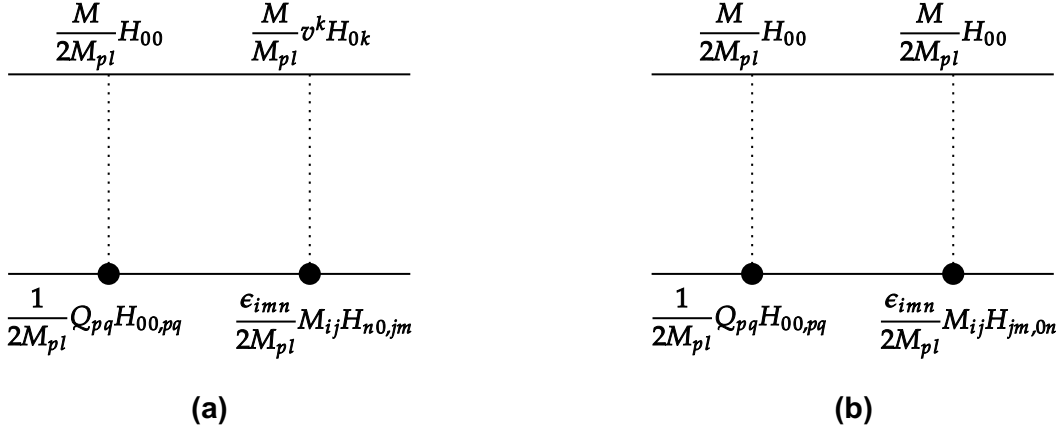

**Figure 10.** Leading Order of $Q$ and $M$ Insertions.

Due to the nature of the dissipative effects, which is asymmetric in time, we have to use the in-in formalism [22, 23] to account for the causal behavior of the system correctly. At the leading order, with $Q$ and $M$ insertions, we have two diagrams:

The sum of both diagrams in figure 10 gives[6]

$$S = \int dt dt' \frac{M^2}{(4M_{pl})^2} \epsilon_{imn} \langle Q_{pq} M_{ij} \rangle \left( \frac{1}{4\pi r} \left( \partial_p \partial_q \partial_n \partial_0 \frac{1}{4\pi r} \right) + 2 \left( \partial_p \partial_q \frac{1}{4\pi r} \right) \left( \partial_0 \partial_n \frac{1}{4\pi r} \right) + v_n q_{pq} q_{jm} \right). \tag{4.19}$$

Assuming the spherical symmetry, only the last term gives us a non-zero contribution. Thus, we are left with[7]

$$S = \int dt dt' \frac{M^2}{(4M_{pl})^2} \epsilon_{imn} \left\langle Q^a_{pq} M^b_{ij} \right\rangle v_n q^a_{pq} q^b_{jm}, \tag{4.20}$$

writing the $\langle QM \rangle$ correlator as

$$\left\langle Q^a_{pq}(t) M^b_{ij}(t') \right\rangle \equiv D^{ab}_{pqij}(t,t'), \tag{4.21}$$

where we use the following basis

$$D^{ab}(t,t') = \begin{pmatrix} 0 & -iD_{adv} \\ -iD_{ret} & \frac{1}{2}D_H \end{pmatrix} \tag{4.22}$$

The equations of motion follow from varying the action

$$\left. \frac{\delta S}{\delta q_-} \right|_{q_-=0, q_+=q(x), J=0} = 0. \tag{4.23}$$

---

[5]Latin indices represent the local frame, where two frames are related by the vierbein as $e^\mu_a e^\nu_b \eta^{ab} = g^{\mu\nu}$.
[6]We introduce $q_{ij} = \partial_i \partial_j \frac{1}{4\pi r}$.
[7]Where $a$ and $b$ are closed-time-path functional indices for the in-in formalism.

– 12 –

Thus, only linear terms in $q_-$ remain and we find

$$F_l = \frac{M^2}{(4M_{pl})^2}\epsilon_{imn}v_n(-2i)\int dt' q_{pq}(t')D^{ret}_{pqim}(t-t')(\partial_l q_{jm}(t)), \qquad (4.24)$$

where we have used $D_{ret}(t,t') = D_{adv}(t',t)$. At this point, we do not know the exact form of the $\langle QM \rangle$ correlator, but we can extract it from the absorption cross section.[8] Using our assumption of spherical symmetry, we write the response of $E$ and $B$ parts in the following form,

$$\begin{aligned}&A^{E+(B+)}_{pqij}(w) = \\ &\int dt e^{-iwt} <0|Q_{pq}(t)M_{ij}(0)|0> = A^{E_+(B+)}(w)\left(\frac{2}{3}\delta_{pq}\delta_{ij} - \delta_{pi}\delta_{qj} - \delta_{pj}\delta_q\right).\end{aligned} \qquad (4.25)$$

If we perform Fourier transformation of $D^{ret}(t-t')$ in our expression for force, then the imaginary part in frequency corresponds to the dissipative part of the force. Since we already have the expression for an imaginary part of $D_{ret}(w)$, and all other parts are real, we can obtain the dissipative part of the force.[9]

$$\begin{aligned}F^{dis}_l &= \frac{M^2(-2i)\epsilon_{imn}v_n}{(4M_{pl})^2} \\ &\times \int dt' \int [dw] e^{-w(t-t')}\frac{\sigma_{abs}}{2w^3}M^2_{pl}\left(\frac{2}{3}\delta_{pq}\delta_{ij} - \delta_{pi}\delta_{qj} - \delta_{pj}\delta_q\right)q_{pq}(t')D^{ret}_{pqim}(t-t')(\partial_l q_{jm}(t)).\end{aligned} \qquad (4.26)$$

As the dissipative part should be odd in powers of $w$ and assuming the correlator has no long-time tails, $\sigma_{abs}$ should be even in positive powers of $w$. Therefore, $\sigma_{abs}$ should start with $w^4$.

Finally, since $\sigma_{abs}$ must scale as length square and the only relevant scale for a compact object is its radius, $\sigma_{abs}$ will have the following form

$$\sigma_{abs}(w) = A_4 R^6 w^4 + A_6 R^8 w^6 + \ldots \qquad (4.27)$$

Inserting the leading order term into the force, we obtain

$$\begin{aligned}F^{dis}_l &= -i\frac{M^2\epsilon_{imn}v_n}{16M^2_{pl}} \\ &\times \int dt' \int [dw] e^{-w(t-t')}A_4 R^6 w^4\left(\frac{2}{3}\delta_{pq}\delta_{ij} - \delta_{pi}\delta_{qj} - \delta_{pj}\delta_q\right)q_{pq}(t')D^{ret}_{pqim}(t-t')(\partial_l q_{jm}(t)).\end{aligned} \qquad (4.28)$$

Performing the integration and summation over indices, we get the dissipative force $F_b$ on compact object $b$, due to its partner's mass $M_a$.

$$\vec{F}^{dis}_b = -9G\frac{M_a^2 A_4^b R_b^6}{\pi r^{10}}(\mathbf{x}\cdot\mathbf{v})(\mathbf{x}\times\mathbf{v}), \qquad (4.29)$$

---

[8]Once we have a phenomenology, we will have the absorption cross section to match.

[9]We make the same spectral decomposition for $D^{ret}_{pqij}$ as we have made for $A^+_{pqij}$.

– 13 –

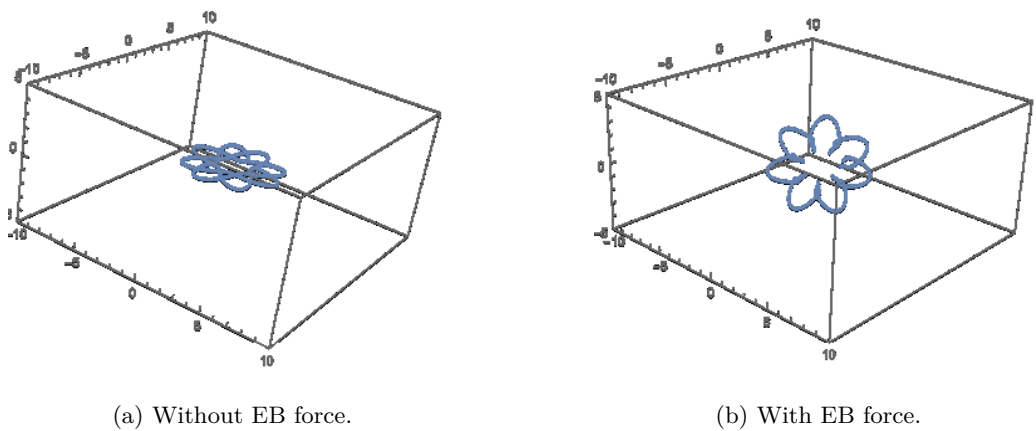

(a) Without EB force.  (b) With EB force.

**Figure 11**. Orbit.

where $R_b$ is the radius of a compact object $b$ and $A_4^b$ its corresponding value from (4.27), which should be the function of the equation of the state for the compact object $b$ and can be fixed for certain types of stars. To see how the new type of force affects the orbit, we numerically solve the equations of motion with and without $F^{dis}$. Figure 11 depicts the orbit aberration, but orbit is still 'closed'. This abberation might appear already within the bandwith of our detectors due to the fact that the finite size effects 'turn on' at later stages of the inspiral.

To compare to the dissipative force, due to the parity conserving effects, we can write the same type of diagrams as in figure 10 but with both $Q$ and $Q$ insertions, which provides the corresponding dissipative force

$$\vec{F}^{dis}_{(GR)b} = -9G\frac{M_a^2 A_4^b R_b^6}{32\pi r^8}\Big(\mathbf{v} + 2\frac{\mathbf{x}\cdot\mathbf{v}}{r^2}\mathbf{x}\Big). \qquad (4.30)$$

We can fix $A_4^b$ in the case of black holes. At leading order in the derivative expansion, the absorptive cross section for a graviton scattering on a black hole is given by [35]

$$\sigma_{abs} = \frac{4\pi r_s^6 \omega^4}{45}, \qquad (4.31)$$

so that $R_b = 2GM$ and $A_4^b = \frac{4\pi}{45}$, which enables us to write dissipative force for the black hole

$$\vec{F}^{dis}_{(GR)b} = M_a^2 M_b^6 \frac{8G^7}{5r^8}\Big(\mathbf{v} + 2\frac{\mathbf{x}\cdot\mathbf{v}}{r^2}\mathbf{x}\Big). \qquad (4.32)$$

Even though we cannot fix the $A_4^b$ in the same way, without additional input about the parity violating compact object, we can see that dissipative force for the parity violating compact object comes at the same PN order, as its parity conserving analog.

## 5 Waveforms

To qualitatively analyze new effects on the dephasing of the waveform, we consider a simplified example with the mass quadrupole radiation. We consider circular motion with

– 14 –

radius $R$ and choose the $(x, y, z)$ frame so that the orbit lies in the $(x, y)$ plane, and is given by

$$\begin{aligned} x(t) &= R\cos(\omega_s t), \\ y(t) &= R\sin(\omega_s t), \\ z(t) &= 0. \end{aligned} \qquad (5.1)$$

Considering eq. (4.5), for our case, we have the following non-zero components

$$Q_{13} + Q_{31} = \frac{CM}{4M_{pl}^2} \frac{1}{4\pi r^3} (v_2 n_1 n_1 - v_1 n_1 n_2) = \frac{CM}{4M_{pl}^2} \frac{1}{4\pi r^3} R\omega_s \cos(\omega_s t), \qquad (5.2)$$

$$Q_{23} + Q_{32} = \frac{CM}{4M_{pl}^2} \frac{1}{4\pi r^3} (v_2 n_2 n_1 - v_1 n_2 n_2) = \frac{CM}{4M_{pl}^2} \frac{1}{4\pi r^3} R\omega_s \sin(\omega_s t). \qquad (5.3)$$

Typical second mass moment is $M^{ij} = \mu x^i(t) x^j(t)$ and for our case, non-zero elements are

$$M_{11} = \mu R^2 \frac{1 - \cos(2\omega_s t)}{2}, \qquad (5.4)$$

$$M_{22} = \mu R^2 \frac{1 + \cos(2\omega_s t)}{2}, \qquad (5.5)$$

$$M_{12} = -\frac{1}{2}\mu R^2 \sin(2\omega_s t). \qquad (5.6)$$

The total mass quadrupole then is

$$M_{ij}^{\text{tot}} = M_{ij} + Q_{ij}.$$

In the general case, mass quadrupole radiation in the Transverse-Traceless Gauge has the following contribution to plus and cross polarizations of gravitational waves

$$\begin{aligned} h_+(t; \theta, \phi) = \frac{1}{r_0} \frac{G}{c^4} \Big( & \ddot{M}_{11}^{\text{tot}}(\cos^2\phi - \sin^2\phi \cos^2\theta) \\ & + \ddot{M}_{22}^{\text{tot}}(\sin^2\phi - \cos^2\phi \cos^2\theta) \\ & - \ddot{M}_{33}^{\text{tot}} \sin^2\theta \\ & - \ddot{M}_{12}^{\text{tot}} \sin 2\phi (1 + \cos^2\theta) \\ & + \ddot{M}_{13}^{\text{tot}} \sin\phi \sin 2\theta \\ & + \ddot{M}_{23}^{\text{tot}} \cos\phi \sin 2\theta \Big), \end{aligned} \qquad (5.7)$$

$$\begin{aligned} h_\times(t; \theta, \phi) = \frac{1}{r_0} \frac{G}{c^4} \Big( & (\ddot{M}_{11}^{\text{tot}} - \ddot{M}_{22}^{\text{tot}}) \sin 2\phi \cos\theta \\ & + 2\ddot{M}_{12}^{\text{tot}} \cos 2\phi \cos\theta \\ & - 2\ddot{M}_{13}^{\text{tot}} \cos\phi \sin\theta \\ & + 2\ddot{M}_{23}^{\text{tot}} \sin\phi \sin\theta \Big). \end{aligned} \qquad (5.8)$$

Where $r_0$ is the distance to our detectors and $\theta$ is the angle between the normal from the inspiral plane and direction to detectors, which effectively stays the same during the inspiral.

Since the orbit is circular, it will be invariant under changes of $\phi$ and we can pick it to be zero. For algebraic simplicity, we assume that inspiral happens in the aligned plane to our detectors, so we set $\theta = \pi/2$. Therefore, with our settings, equations (5.7) and (5.8) for gravitational waves obtain the simplified form

$$h_+(t) = \frac{1}{r_0}\frac{G}{c^4}\ddot{M}_{11}, \tag{5.9}$$

$$h_\times(t) = -\frac{1}{r_0}\frac{G}{c^4}\left(\ddot{Q}_{13} + \ddot{Q}_{31}\right) \tag{5.10}$$

We see that in such a scenario, when $\theta = \pi/2$, we can have cross polarization only due to the parity violating effects. Using (5.2) and (5.4) we can obtain the explicit expression for the gravitational waves,

$$h_+(t) = \frac{2}{r_0}\frac{G\mu w_s^2 R^2}{c^4}\cos(2w_s t), \tag{5.11}$$

$$h_\times(t) = \frac{1}{r_0}\frac{G^2 M \omega_s^3}{c^5}\frac{C}{4R^2}\cos(w_s t), \tag{5.12}$$

where for the last equation we have used the fact that distance between compact objects is twice the radius $r = 2R$ and note that we seemingly have $G^2$, but as we will see $C \sim M_{pl}^2 R^5 \sim R^5/G$. We see that contribution from the parity violating effects to the gravitational waves come at +1PN compared to the parity conservation effects. To analyze the evolution, we need to consider the power loss due to the radiation of gravitational-waves, which will induce change in the orbital frequency. The angular distribution of the radiated power, in the quadrupole approximation, is obtained by

$$\left(\frac{dP}{d\Omega}\right)_{quad} = \frac{r^2 c^3}{16\pi G}\langle \dot{h}_+^2 + \dot{h}_\times^2 \rangle. \tag{5.13}$$

Using power loss, we estimate rate of change in the frequency at the leading order,

$$\dot{w}_{gw} = \frac{12}{5}2^{1/3}\left(\frac{GM}{c^3}\right)^{5/3}w_{gw}^{11/3}. \tag{5.14}$$

Where $w_{gw} = 2w_s$ and integrating its time-evolution equation we see that it formally diverges. But the divergence is cut off by the fact that, when their separation becomes smaller than a cirtical distance, the two compact objects merge. Because we consider change in frequency, we have to take that into account for coordinates, which do not have $w_s t$ as their argument. Instead, we have $x(t) = R\cos(\Phi(t)/2)$ and $y(t) = R\cos(\Phi(t)/2)$, where we define

$$\Phi(t) = 2\int_{t_0}^t dt' w_s(t'). \tag{5.15}$$

The unknown coefficient $C$ should scale the same as $C_E$ and $C_B$. Because while $C_E$ ($C_B$) provide as quadrupolar mass (current) moment response to the external quadrupolar electro(magneto) gravitational field, C should provide the mass response to gravitomagnetic field and vice-versa. Thus, we write it as

$$C = M_{pl}^2 R^5 k_2^{EB}, \tag{5.16}$$

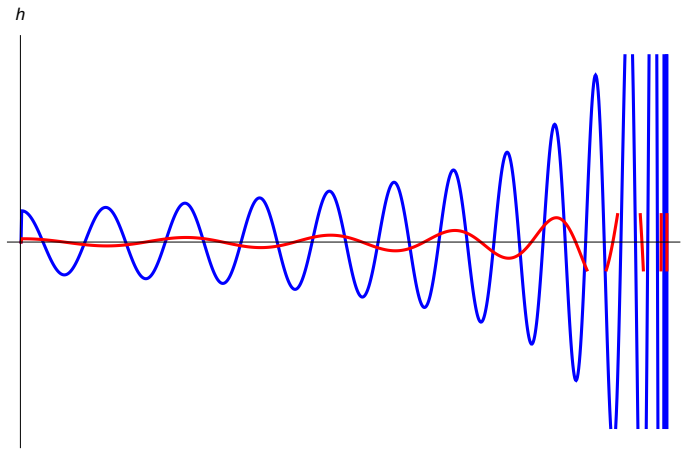

**Figure 12**. Waveform for $h_+$ (blue) and due to the parity violating effects, $h_\times$ (red). For compact objects of similar masses and $k_2^{EB}$ the order of one.

where $k_2^{EB}$ will be the new, corresponding Love number, which should be a function of the equation of state of the compact object. To numerically analyze different scenarios, with and without new effects, we take $k_2^{EB}$ to be the order of one, which is typical for $k_2^E$.

We see that the new effects become relevant on the later stages of the inspiral, which was expected due to the finite size effects. Another interesting feature is swapping minimums for each cycle, which we could predict from eq. (5.11), where frequence for the parity violating effects is twice less. It is remarkable that in the given scenario, we would not expect cross polarization in gravitational waves, unless there is finite size parity violating effects. This is another unique signature in gravitational waves, due to the parity violating effects.

## 6 Conclusion

We derived the finite size effects for spinning binary inspirals within the EFT formalism, and we obtained the leading order potentials and radiating moments that agree with previous results. We generated the current quadrupole moment, induced by the gravitomagnetic effects, which is a new result.

We generalized the EFT approach to include the finite size parity violating effects and discovered some unique signatures. In particular, the Post Newtonian order of the mass and current moments, sourced from the parity violating finite size effects, is swapped. Also, generalizing to the spinning case, a new type of potential, $\alpha \mathbf{S} \cdot \mathbf{r}$ was found. Finally, we considered parity violating absorption effects, which lead to a force, normal to the orbital plane.

It would be interesting to consider models of parity violating compact objects to match for $C$. This can be accomplished by calculating the induced mass moment $Q_E$ due to a background gravitomagnetic $B$ field.

### Acknowledgments

I thank Walter Goldberger and Ira Rothstein, for their helpful comments. I am grateful to Ira Rothstein, for encouragement and invaluable discussions during this work, which elucidated many aspects.